# Dirac Electrons in Molecular Solids

**Hidetoshi Fukuyama**

Faculty of Science and Research Institute for Science and Technology

Tokyo University of Science, Tokyo 162-8601

**Akito Kobayashi**

Institute for Advanced Research, Nagoya University, Nagoya 464-8602

**Yoshikazu Suzumura**

Department of Physics, Nagoya University, Nagoya 464-8602

fukuyama@rs.kagu.tus.ac.jp

**Abstract**. Electrons in solids are characterized by the energy bands, which indicate that electrons are considered to be "elementary particles" with specific effective masses and g-factors reflecting features of each solid. There are cases where these particles obey dispersion relationship similar to those of Dirac electrons. Examples include graphite and bismuth both of which are known for many years, together with graphene, a single layer of graphite, recently addressed intensively after its realization. Another recent example is a molecular solid, $\alpha$-ET$_2$I$_3$, which is described by an equation similar to Weyl equation with massless Dirac cones but the coordinate axis is tilted because of the location of cones at off-symmetry points. Orbital susceptibility of such Dirac electrons in graphite and bismuth has been known to have striking features not present in ordinary band electrons but resulting from the inter-band matrix elements of magnetic field. Results of theoretical studies on not only orbital susceptibility but also Hall effect of such Dirac electrons in molecular solids with tilting are introduced in this paper.

## 1. Introduction

Energy bands of electrons in solids have diversity reflecting particular features of solids and disclose specific energy-momentum relationship, which can be viewed as particular "elementary particles" at least locally in momentum space with effective masses and g-factors generally with spatial anisotropy. There are cases where these elementary particles obey equations similar to Dirac electrons. In this context graphite[1] and bismuth (Bi)[2, 3] have been known for many years; 2x2 Weyl equation for the former and 4x4 Dirac equation for the latter. Main physical interests in these systems had been the orbital susceptibility in weak magnetic field especially in bismuth (Bi) [4]. The difficulty with Landau-Peierls formula based on the pioneering work by Peierls [5], which takes only intra-band contribution of the effects of magnetic field on Bloch electrons, had long been noticed in view of experimental findings and roles of inter-band effects have been recognized [6]. Accordingly since the first derivation of orbital susceptibility by Wilson [7] there have been many efforts to derive exact formula of orbital susceptibility of Bloch electrons [8]. Theoretical understanding of experimentally observed anomalous orbital susceptibility has been achieved for graphite[9] and Bi[10] by taking these inter-band contributions of vector potential based on k-p Hamiltonian (Luttinger–Kohn representation) [11] which is suited to explore such intricate effects in a precise and transparent way. A simple and exact formula [12] for the orbital susceptibility for Bloch electrons has been derived based on this k-p Hamiltonian.

Besides orbital susceptibility, which is thermodynamic, there is general interest in such inter-band contribution on transport properties, typically weak field Hall conductivity and then Hall effect [13, 14]. These have been explicitly studied for the recent example of Dirac electrons in graphene [15, 16], a single layer of graphite. Based on the proper scheme [17] in extracting correctly (in a gauge-invariant way) the contributions in the linear order of vector potential in the Kubo formula Hall conductivity has been evaluated [18] together with conductivity and orbital susceptibility at absolute zero as a function of chemical potential under the assumption of energy independent electron damping (lifetime). The results indicate that Hall conductivity vanishes and change signs at the crossing of Dirac cone. This feature at the crossing energy is very natural in view of the symmetry between "electrons" and "holes". This fact is, however, completely opposite to the conventional understanding based on the band electrons, since the effective carrier density would be vanishing if the Fermi energy is located at the crossing.

## 2. Experimental puzzles in molecular solids, $\alpha$-ET$_2$I$_3$

There had been various puzzles in the experimental results in $\alpha$-ET$_2$I$_3$ [19], which is one of polytypes of charge–transfer molecular solids consisting of ET molecules, ET$_2$X, with X being anions. At ambient pressure the resistivity of $\alpha$-ET$_2$I$_3$ shows a sharp increase at around 150K as the temperature is lowered [19]. This metal-insulator transition is suppressed by external pressure and vanishes completely at around 20 GPa leading to almost temperature independent resistivity below room temperature. The measurement of Hall effect in this pressure region of temperature-independent resistivity discloses that the effective carrier number decreases very strongly (several orders of magnitude between room temperature and few degree K) as temperature is lowered, whereas the mobility increases accordingly. Even more, there are cases where Hall constant changes sign at low temperature very sharply [20], which is sample-dependent.

## 3. Electronic states of molecular solids, $\alpha$-ET$_2$I$_3$: tilted Weyl particles

It is now established that electronic states of molecular solids are, in general, well described by the tight-binding approximation based on molecular orbitals [21]. Molecular solids of charge transfer type with particular ratio of 2:1 between cations and anions, which lead to the quarter-filling of holes, have a great diversity in electronic properties. Microscopic and systematic theoretical understanding of the diversity is now possible by the extended Hubbard model which is based on the tight-binding approximation in terms of molecular orbitals for the band structure together with Coulomb interactions U on the same molecules and V between them [22,23]. The characteristic electronic states of quarter-filled band are due to coordinated effects of dimerization and frustration (from square to triangular) of lattice structures resulting in two typical insulating states, dimer Mott and charge ordering [23]. Typical example of the former is the κ type which shows competition between antiferromagnetic Mott insulator and superconductivity [24] and the realization of spin liquid [25] for the first time, while α-ET$_2$I$_3$ is the example of the latter[26].

For α-ET$_2$I$_3,$ using the transfer energy experimentally obtained under uniaxial strain [27], the gapless behaviour in the density of states has been disclosed [28]. Further, detailed analysis of such band structure within the tight-binding approximation [29] and later by first-principles calculations [30, 31 ] have led to the remarkable finding of the possible accidental degeneracy of bands (Dirac cones). This finding has led to the breakthrough of microscopic understanding of mysteries of this system.

Since Dirac cones in this case are located at two points (with inversion symmetry) in the Brillouin zone away from symmetry points in contrast to graphenes, the band energy is described by the following equation similar to Weyl equation but with diagonal component linear in wave vector, k, measured from the crossing point leading to "tilting" of cones[32].

$$H = \sum_{\rho=0,x,y,z} \mathbf{k} \cdot \mathbf{v}_\rho \sigma_\rho \quad (1)$$

Here $\hbar = h/2\pi$ taken as unit with h being the Plank constant and $\sigma_\rho$ are 2x2 matrices with $\sigma_0$ being unit matrix and the Pauli matrix for $\rho$=x, y, z, respectively. It turns out that velocity $\mathbf{v}_\rho$ are given by $\mathbf{v}_0 = (v_0, 0)$, $\mathbf{v}_x = (v_x, 0)$, $\mathbf{v}_y = 0$, $\mathbf{v}_z = (0, v_z)$ and that $v_x$ is close to $v_z$ and then $v_x = v_z$ is assumed in the explicit calculations. The existence of finite $v_0$ represents tilting, which is characterized by the dimensionless parameter, $\alpha = v_0/v_c$. Here $v_c$ is defined by $v_c = v_x = v_z$ [32]. It turns out that $\alpha \sim 0.8$ in the case of $\alpha$-ET$_2$I$_3$.

## 4. Effects of tilting on orbital susceptibility and Hall conductivity

The effects of tilting on conductivity and Hall conductivity have been studied [33] in terms of proper formulas applied to the Hamiltonian, eq. (1). The results are shown in Fig.1 and 2 at absolute zero as a function of chemical potential under the assumption of energy independent electron damping (lifetime). Here $X = \mu/\Gamma_0$, with $\mu$ and $\Gamma_0$ being the chemical potential and damping, respectively. In the case of conductivity, there are two different types of contributions, i.e. $\sigma_{xx} = \sigma_{xx}^a + \sigma_{xx}^b$, where both $\sigma_{xx}^a$ and $\sigma_{xx}^b$ have contributions from only at the Fermi energy (on-shell contributions) but have different dependences on $\alpha$. (There is a finite anisotropy in the conductivity but it is small as is expected from the weak $\alpha$ dependences of $\sigma_{xx}$, especially near the crossing, X=0, as seen in the inset of Fig.1.) Similarly the Hall conductivity has two contributions, $\sigma_{xy}= \sigma_{xy}^a + \sigma_{xy}^b$. It is to be noted, however, that there is a qualitative difference from conductivity; $\sigma_{xy}^a$ contains all contributions below the Fermi energy whereas $\sigma_{xy}^b$ has only on-shell contributions. In Fig.3 is shown the result for orbital susceptibility, in which case tilting affects only the over-all prefactor. In all these singular behaviour toward $\alpha =1$ is due to the fact that the velocity in one direction is vanishing in this limit.

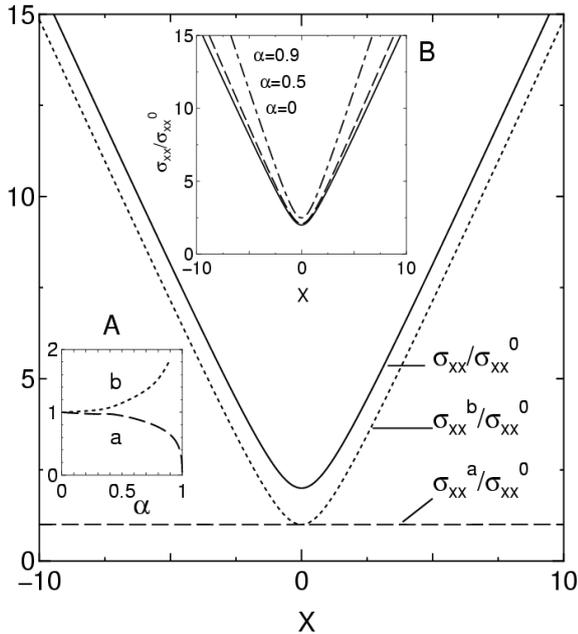
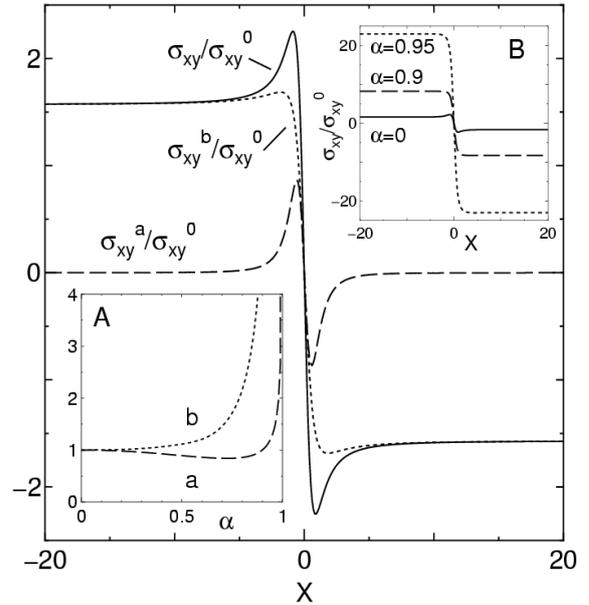

Fig. 1. The conductivity $\sigma_{xx}/\sigma_{xx}^0$ (the solid line), $\sigma_{xx}^a/\sigma_{xx}^0$ (the dashed line) and $\sigma_{xx}^b/\sigma_{xx}^0$ (the dotted line) for $\alpha=0$, where $\sigma_{xx}^0 = e^2/2\pi^2$. The inset A shows $\alpha$-dependences of enhancement factors, $\sigma_{xx}^a(\alpha)/\sigma_{xx}^a(\alpha=0)$ (the dashed line) and $\sigma_{xx}^b(\alpha)/\sigma_{xx}^b(\alpha=0)$ (the dotted line). The inset B shows X-dependences of total $\sigma_{xx}$ for several degrees of tilting, $\alpha=0$ (the solid line), 0.5 (the dashed line), and 0.9 (the dot-dashed line). [33]

Fig. 2. The Hall conductivity $\sigma_{xy}/\sigma_{xy}^0$ (the solid line), $\sigma_{xy}^a/\sigma_{xy}^0$ (the dashed line) and $\sigma_{xy}^b/\sigma_{xy}^0$ (the dotted line) in the absence of tilting, where $\sigma_{xy}^0 = e^3 v_c^2 H/4\pi^2 c\Gamma_0^2$. The inset A shows $\alpha$-dependences of enhancement factors, $\sigma_{xy}^a(\alpha)/\sigma_{xy}^a(\alpha=0)$ (the dashed line) and $\sigma_{xy}^b(\alpha)/\sigma_{xy}^b(\alpha=0)$ (the dotted line). The inset B shows X-dependences of total $\sigma_{xy}$ for several degrees of tilting. [33]

It is seen that finite tilting affects both orbital susceptibility and Hall conductivity quantitatively but qualitative features are unchanged except close to $\alpha=1$. It is to be noted that the present assumption of energy-independent damping of electrons leads to increase of conductivity as a function of energy. This result will lead to the increase of conductivity as a function of temperature in the case where the Fermi energy is located at the crossing energy at the absolute zero. This is not consistent with the experimental results. This will imply that the damping is increasing as a function of temperature (energy) almost linearly. If this is the case, the Hall coefficient will be strongly suppressed as the temperature is increased as seen in the experiment, although it is a future problem to explore these dependences on energy at absolute zero and on temperature at fixed carrier density.

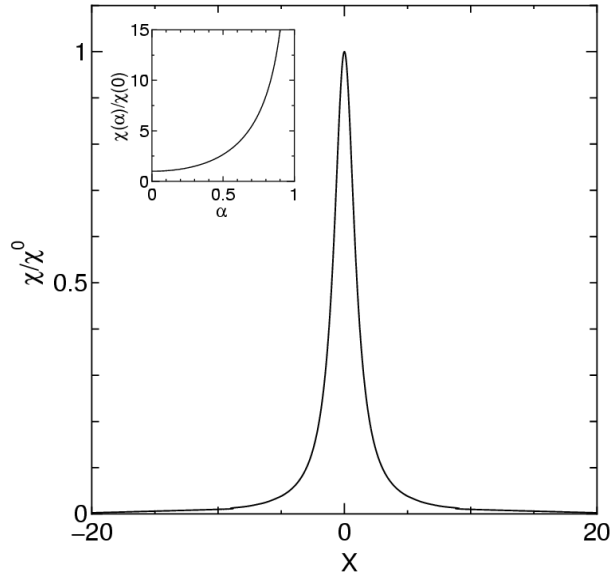

Fig. 3. The orbital susceptibility $\chi/\chi^0$ as the function of X in the absence of tilting, where $\chi^0=- e^2 v_c^2/3 \pi^2 c^2 \Gamma_0$. The inset shows $\alpha$−dependence of enhancement factor $\chi (\alpha) / \chi (\alpha=0)$. [33]

Regarding the sharp change of sign in Hall coefficient at low temperatures, it has been indicated [33] that the possible small amount of deficiency of anions ($I_3^-$) of the order of only ppm together with asymmetry of energy dependences of density of states relative to the crossing energy can be the cause.

## 5. A possible Berezinski-Kosterlitz-Thouless phase transition under strong magnetic field

So far the orbital susceptibility and Hall conductivity have been studied in weak magnetic field. Under strong magnetic field of the order of 10 Tesla applied perpendicularly, resistivity measurement indicates interesting temperature dependences[19], which shows the existence of two characteristic temperatures, $T_0$ and $T_1$, for its sharp change. It has been proposed [34] that these characteristic temperatures will correspond to the crossover temperature of the formation of valley-order similar to the XY ferromagnetic order and the Berezinski-Kosterlitz-Thouless (BKT) transition at lower temperature as will be briefly explained below.

Magnetic field applied perpendicularly to the plane leads to the quantized Landau levels. Note that there exists a particular Landau level at the crossing energy (N=0) independent of the strength of magnetic field. The finite tilting does not affect this qualitatively [35] and the energy level separation between N=0 and N=1 is of the order of 50K at H=10T.

The spin Zeeman splitting is much smaller than the Landau level splitting since the g-factor is g~2. Because there are two Dirac cones in the Brillouin zone, these states (valleys) are doubly degenerate (characterized by the pseudo-spin) if two valleys are independent. Once mutual Coulomb interactions are taken into account, electrons in different valleys interact and get scattered among them. In order to treat such many-body effects in the presence of Landau quantization, ordinary Landau wave functions, which are plane wave and then infinitely spread in one direction, is not convenient and then orthonormal basis sets, which are localized in space similar to "Wannier functions", introduced some time ago are employed.

It turns out that finite tilting plays crucial roles by introducing processes of mixing different valleys. If the chemical potential is fixed to the crossing energy and this mixing is strong enough compared with the spin Zeeman splitting, some kind of valley order state described as ferromagnetism of pseudo-spins is possible at $T=T_0$ in the mean-filed approximation. The order parameters in this case are complex and their phases represent degrees of freedom leading to vortices which can have BKT phase transition at lower temperature $T=T_1$. Theoretical analysis concludes $T_1/T_0 \sim 1/4$. If $T_0$ and $T_1$ are associated with the two step changes of resistivity in the experiment [19] this ratio between $T_0$ and $T_1$ is close to experimental results. A phase transition similar to BKT had been indicated in resistivity measurement of graphene [36] at much lower temperatures. Since tilting is absent in graphene, some mechanism other than the present tilting-driven inter-valley mixing by Coulomb interactions, *e. g.* an electron-lattice interaction [37], will be operating in graphene.

**Acknowledgment**

One of authors (HF) dedicates hearty gratitude to late Professor Ryogo Kubo for his guidance in very early stage based on the deep insight into inter-band effects of magnetic field on Bloch electrons. Authors are grateful to Mark Goerbig for collaborations for the subject in Sec.5.

**References**

[1]     P. R. Wallace, Phys. Rev. 71, 622 (1947)
[2]     M. H. Cohen and E. I. Blount, Phil. Mag. 5, 115 (1960)
[3]     P. A. Wolff, J. Phys. Chem. Solids 25, 1057 (1964).
[4]     D. Schoenberg and M. Z. Uddin, Proc. Roy. Soc. A156, 687,701 (1936).
[5]     R. Peierls, Z. Phys. 80, 763 (1933).
[6]     E. N. Adams, Phys. Rev. 89, 633 (1953).
[7]     A. H. Wilson, Proc. Camb. Phil. Soc. 49, 292 (1953).
[8]     For example, G. H. Wannier and U. N. Upadyaya, Phys. Rev. 136, A803 (1964); R. Kubo. Proc. Intern. Conf. Magnetism, Notthingam 1964 (The Institute of Physics and Physical Society) p.35, and references therein.
[9]    J. W. McClure, Phys. Rev. 104, 666 (1956).
[10]   H. Fukuyama and R. Kubo, J. Phys. Soc. Jpn. 28, 570 (1970).
[11]   J. M. Luttinger and W. Kohn, Phys. Rev. 97, 869 (1955).
[12]   H. Fukuyama, Prog. Theor. Phys. 45, 704 (1971).
[13]   R. Kubo and H. Fukuyama, Proc. of the "10$^{th}$ International Conference on the Physics of Semiconductors" (Ed. E. P. Keller and J. C. Hensel and F. Stern, Published by the United State Atomic Energy Commission, 1970).
[14]   H. Fukuyama, Ann. Phys. (Leipzig) 15, 520 (2006).
[15]   K. S. Novoselov, A. K. Geim, S. V. Morozov, D. Jiang, M. I. Katsnelson, I. V. Grigorieva, S. V. Dubonos, and A. A. Firsov: Nature 438, 197 (2005).
[16]   T. Ando: J. Phys. Soc. Jpn. 74, 777 (2005).
[17]   H. Fukuyama, H. Ebisawa and Y. Wada, Prog. Theor. Phys. 42, 494 (1969).
[18]   H. Fukuyama, J. Phys. Soc. Jpn. 76, 043711 (2007).
[19]   N. Tajima, S. Sugawara, M. Tamura, Y. Nishio and K. Kajita, J.Phys. Soc. Jpn. 75, 0510101 (2005).
[20]   N. Tajima, S. Sugawara, M. Tamura, R. Kato, Y. Nishio and K. Kajita, private communications.
[21]   T. Mori, A. Kobayashi, T. Sasaki, H. Kobayashi, G. Saito and H. Inokuchi, Chem. Lett. 13, 957 (1984) and references therein.

[22] H. Kino and H. Fukuyama, J.Phys. Soc. Jpn. 65, 2157 (1996).
[23] H. Seo, C. Hotta and H. Fukuyama, Chem. Rev. 104, 5005 (2004).
[24] K. Kanoda, J. Phys. Soc. Jpn. 75, 051007 (2005).
[25] K. Kanoda and R. Kato, Annu. Rev. of Condens. Matter Phys. 2011. 2:167-88
[26] H. Seo, J. Phys. Soc. Jpn. 69, 805 (2000).
[27] R. Kondo, S. Kagoshima, and J. Harada, Rev. Sci. Instrum. 76, 090302 (2005).
[28] A. Kobayashi, S. Katayama, K. Noguchi, and Y. Suzumura, J. Phys. Soc. Jpn 73, 3135 (2004).
[29] S. Katayama, A. Kobayashi, and Y. Suzumura, J. Phys. Soc. Jpn 75, 054705 (2006).
[30] H. Kino and T. Miyazaki, J. Phys. Soc. Jpn 75, 034704 (2006).
[31] S. Ishibashi, T. Tamura, M. Kohyama, and K. Terasaki, J. Phys. Soc. Jpn 75, 015005 (2006).
[32] A. Kobayashi, S. Katayama, Y. Suzumura, and H. Fukuyama, J. Phys. Soc. Jpn 76, 034711 (2007).
[33] A. Kobayashi, Y. Suzumura, and H. Fukuyama, J. Phys. Soc. Jpn 77, 064718 (2008).
[34] A. Kobayashi, Y. Suzumura, H. Fukuyama, and M. O. Goerbig, J. Phys. Soc. Jpn 78, 11471, (2009).
[35] T. Morinari, T. Himura, and T. Tohyama, J. Phys. Soc. Jpn 78, 023704 (2009).
[36] J. G. Checkelsky, L. Li, and N. P. Ong, Phys. Rev. B 79, 115434 (2009).
[37] K. Nomura, S. Ryu, and D-H Lee : Phys. Rev. Lett. 103 (2009) 216801.